\documentclass[10pt,twocolumn,aps,floatfix,pre,
                superscriptaddress,longbibliography]{revtex4-1}
\pdfoutput=1 

\makeatletter
\renewcommand{\section}{
  \@startsection{section}{1}{0pt}
  {1.5ex}   
  {0.8ex}   
  {\centering\normalfont\bfseries}
}
\makeatother
\usepackage{amsmath,amssymb, eucal,graphicx,float,epstopdf,bm}
\newcommand{\eq}[1]{Eq.~(\ref{eq:#1})}
\newcommand{\fig}[1]{Fig.~\ref{fig:#1}}
\newcommand{\tbl}[1]{Table~(\ref{tbl:#1})}
\newcommand{\sectn}[1]{Sec.~(\ref{sec:#1})}

\usepackage{enumitem}

\usepackage[colorlinks=true, urlcolor=blue, anchorcolor=blue,
            citecolor=blue, filecolor=blue, linkcolor=blue,
            menucolor=blue]{hyperref}
\usepackage{soul}
\usepackage[normalem]{ulem}

\def\nR{30000}
\def\nRsurf{5000}

\def\aRimp{0.423}
\def\daRimp{0.014}
\def\aVimp{-0.399}
\def\daVimp{0.017}
\def\aCimp{0.363}
\def\daCimp{0.007}
\def\auimp{-0.446}
\def\dauimp{0.042}
\def\ano{-0.168}
\def\dano{0.034}
\def\aTo{-0.692}
\def\daTo{0.035}

\newcommand{\nRimp}{\the\numexpr 4*\nR \relax}

\begin{document}
\title{Impurity dynamics in a zero-temperature gas}
\author{Umesh Kumar}
\email{umesh.kumar@icts.res.in }
\author{Abhishek Dhar}
\email{abhishek.dhar@icts.res.in}
\affiliation{International Centre for Theoretical Sciences,
            Tata Institute of Fundamental Research,
            Bangalore 560089, India}
\author{P. L. Krapivsky}
\email{paulk@bu.edu}
\affiliation{Department of Physics,
            Boston University,
            Boston, MA 02215, USA}
\affiliation{Santa Fe Institute, Santa Fe, NM 87501, USA}

\begin{abstract}
If energy is suddenly released in a localized region of space uniformly filled with identical stationary hard spheres, the outcome is a blast with an asymptotically spherical shock wave separating moving and stationary hard spheres. The radius $R(t)$ of the region filled with the moving spheres grows as $t^{2/(d+2)}$, where $d$ is the spatial dimension. The simplest way to inject energy is to kick a few `impurity' particles. Using hydrodynamics and kinetic theory, we argue that the typical displacement of an impurity scales as $R_{\rm imp} \sim \lambda (R/\lambda)^{(4+3d^2)/(8+3d^2)}$, where $\lambda$ is the mean-free path in the initial state. The number of collisions experienced by each impurity grows as $(R/\lambda)^{(8+2d^2)/(8+3d^2)}$, while its average speed decreases as $t^{-d(8-2d+3d^2)/[(2+d)(8+3d^2)]}$. In $2D$, the predictions for impurity displacement, collision numbers, and speed are $t^{2/5},~t^{2/5}$ and $t^{-2/5}$, respectively. These predictions are in reasonable agreement with the results of molecular dynamics simulations.
\end{abstract}

\maketitle

\section{Introduction}

A localized instantaneous release of energy generates a blast, first investigated in the context of atomic explosions \cite{Taylor50a, Taylor50b,Sedov,vN}. The blast is strong as long as the pressure behind the shock wave greatly exceeds the atmospheric pressure. For the strong blast, Euler equations describing an ideal compressible gas admit an exact solution; see textbooks \cite{Zeldovich,Landau-FM,Barenblatt}. The shock wave eventually weakens, and the problem becomes analytically intractable. However, we look at the idealized case where the pressure in front of the shock is zero, so the blast is infinitely strong throughout the evolution. Infinitely strong blasts mimic supernova explosions and other astrophysical phenomena \cite{McKee1988,Chev17,Vink,Sari21}.

One can examine the infinitely strong blast problem at a microscopic level by considering a hard sphere gas with particles initially at rest, i.e., at zero temperature. How does this system evolve when energy is injected into a localized region, for example, by giving energy to a few particles? An infinite system with initially immobile particles has an additional virtue---it is amenable to direct molecular dynamics simulations since there is only a finite number of moving particles. Hence, one can investigate the behavior in a numerically precise way. This system thus provides a fruitful laboratory for probing the validity of hydrodynamics, explaining the popularity of the setting with initially immobile hard spheres and the number of recent papers studying the blast for this system \cite{AKR,Trizac15,Trizac16,Rajesh21a,Rajesh21b,Rajesh22, blast-1d, blast-1d-PF,blast-splash,kumar2025shock,Rajesh25,blast-2d}. These studies compared the predictions of Euler hydrodynamics with the results from microscopic simulations. Somewhat surprisingly, one finds rather good agreement. However, deviations from the predictions of the Euler equations have been observed. These deviations can be understood by taking dissipative effects into account, such as heat conduction in a {\em core} region around the center of the blast; i.e., one needs to use the full Navier-Stokes equations \cite{blast-1d,blast-1d-PF,blast-2d}.

Previous studies focused on coarse-grained observables such as density, temperature, and pressure. The emergence of the hydrodynamic behavior is questionable, at least at first sight, as the gas outside the shock wave has zero temperature. The success of hydrodynamics comes as a pleasant surprise. Going beyond the scope of conventional hydrodynamics predicting macroscopic characteristics, one can study mesoscopic characteristics such as the self-diffusion phenomenon inside a blast. The present work explores precisely this question. Employing hydrodynamics and kinetic theory arguments, we deduce scaling laws describing the characteristics of the impurity.

In Sec.~\ref{sec:summary}, we describe the model and summarize the main results. In Sec.~\ref{sec:der}, we derive the main results using arguments based on kinetic theory. In Sec.~\ref{sec:PQ}, we argue that the distributions $P(\bm{r},t)$ and $Q(\bm{v},t)$ of the position and velocity of the impurity do not satisfy closed equations. The joint position-velocity distribution $\Pi(\bm{r},\bm{v},t)$ provides a more comprehensive description than the individual distributions $P(\bm{r},t)$ and $Q(\bm{v},t)$ together. The joint position-velocity distribution $\Pi(\bm{r},\bm{v},t)$ satisfies a closed Lorentz-Boltzmann equation. The integro-differential Lorentz-Boltzmann equation in an inhomogeneous and evolving background seems analytically intractable. In Sec.~\ref{sec:test}, we present the results of simulations and their comparison with the analytic predictions. We conclude with a summary in Sec.~\ref{sec:disc}.

\begin{figure}
\centering
\includegraphics[width=\linewidth]{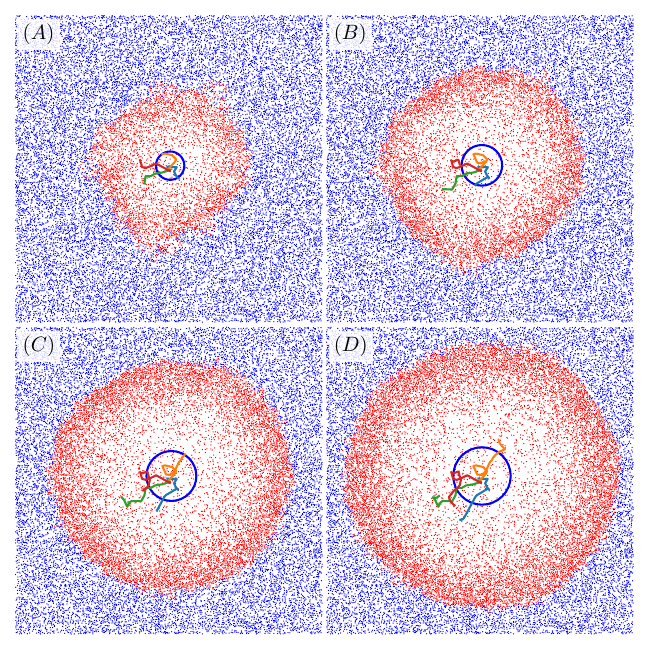}
\caption{Blast in a box of size with $20000$ particles at density $n_{\infty}=0.15$ (packing fraction $\phi=0.118$). $(A), (B), (C)$
and $(D)$ depicts the moving ({\color{red}$\bullet$}) and stationary particles ({\color{blue}$\bullet$}) at times $t=500, 1000, 1500$ and
$2000$ respectively in a single realization. Also shown are the trajectories of four impurity particles. \eq{Rimp_comp} is used to
plot the blue circles in $(A)-(D)$.}
\label{fig:shock-front}
\end{figure}

\section{The model and summary of results}
\label{sec:summary}

Consider a homogeneous gas of hard spheres that are initially at rest, with number density $n_\infty$ and mass density $\rho_\infty$. Each hard sphere has radius $a$ and unit mass ($m=1$); therefore, the mass density and number density are related by $\rho=n$. We perturb the system at time $t=0$ by giving total energy, $E$, to a few particles~\cite{AKR, KRB}. We will refer to these particles as the impurities. The impurities elastically collide with the host spheres, which begin to move and elastically collide with other spheres, generating a growing number of moving spheres. We are primarily interested in the `fate' of the impurities, specifically the asymptotic distributions of their position and velocity. In \fig{shock-front}, we present a simulation of a blast in a two-dimensional hard-disc gas initiated by injecting energy into four impurity particles. The figure also shows the trajectories of these impurity particles against the backdrop of the evolving blast environment.

Following the classic works of Taylor-von-Neumann-Sedov (TvNS) \cite{Taylor50a,Taylor50b,vN,Sedov} on the self-similar solution of the Euler equations, it is known that the moving spheres are separated from quiescent spheres by an asymptotically spherical shock wave advancing as
\begin{equation}\label{R-shock}
R = \left(\frac{E t^2}{A_d n_\infty}\right)^\frac{1}{d+2}
\end{equation}
where $n_\infty$ is the number density (a mass prefactor $m$ has been set to one) of the quiescent spheres and $d$ is the spatial dimension. At first sight, the sub-linear growth $R \sim t^{2/(d+2)}$ is surprising because the shock wave always moves faster than the sonic speed~\cite{Landau-FM}. However, the sonic speed vanishes in the zero-temperature gas, so, albeit the speed of the shock decays as $t^{-d/(d+2)}$, the shock remains infinitely strong throughout the evolution. The position of the shock wave depends only on macroscopic variables, i.e., total energy, $E$, and number density, $n_\infty$. The notation $A_d$ for the amplitude in Eq.~\eqref{R-shock} emphasizes the dependence on the spatial dimension, but $A_d$ also depends on the volume fraction
\begin{equation}\label{phi:def}
\phi = n_\infty\,\Omega_d a^d, 
\quad
\Omega_d = \frac{\pi^\frac{d}{2}}{\Gamma\big(\frac{d}{2}+1\big)}
\end{equation}
occupied by spheres. The analytical expression for amplitudes are generally unknown since even the equation of state for the dense hard-sphere gas is not known in closed form. However,  they can readily be computed numerically. In the $\phi\to 0$ limit, the hard-sphere gas is dilute and ideal. Euler equations for ideal compressible gas behind the shock arising in our infinitely strong blast problem admit an exact solution for arbitrary adiabatic index $\gamma$. We set $\gamma=1+2/d$, the adiabatic index characterizing ideal monoatomic gases \cite{fluid, Beijeren, Landau-K}. In the physically relevant dimensions, the amplitudes are
\begin{equation}
A_2  = 0.351\,935\,906\ldots,
\quad
A_3  = 0.072\,401\,066\ldots
\end{equation}

The work of TvNS showed that the Euler equations, for the hydrodynamic fields for density ($\rho$), velocity ($u$), and temperature ($T$), admitted self-similar scaling forms in terms of the scaling variable $r/R(t)$. For the ideal gas, the scaling functions can be determined analytically in all dimensions~\cite{blast-2d}---we will refer to these as the TvNS solution. However, hard-sphere simulations in Refns.~\cite{blast-1d,blast-1d-PF,blast-2d} have revealed deviations from the solution of the Euler equations for the  compressible gas. They showed that the TvNS predicted growth law in Eq.~\eqref{R-shock} was accurate and the scaling solutions were valid in a large volume of the gas behind the shock front, which we will refer to as the  \emph{bulk}. However, it was found that in a smaller region centered around the blast center, one needs to account for heat conduction. In this {\em core} region, which grows slower than the shock wave, one has to rely on the Navier-Stokes equations~\cite{blast-1d,blast-1d-PF,blast-2d}. While there is no sharp boundary separating the core from the bulk described by Euler equations, one can still talk about the characteristic radius of the core region. The temperature in the core decays with time, yet the ratio of this temperature to the temperature in the bulk of the moving gas behind the shock diverges with time. Thus, the core region is `hot'; we denote its characteristic radius by $H$. This radius grows as \cite{blast-2d}
\begin{equation}
\label{size-dim}
H = \lambda \left(\frac{R}{\lambda}\right)^{h_d},
\qquad
h_d = \frac{4 + 3d^2}{8 + 3 d^2},
\end{equation}
{and $\lambda$ is  the  mean-free path for the quiescent hard sphere gas, given by
\begin{equation}
\lambda = \frac{1}{n_\infty \Omega_{d-1}a^{d-1}}.
\label{eq:mfp}
\end{equation}
We revisit the derivation of this formula in the next section.

Our primary interest in the present work is the `fate' of the impurities --- specifically the asymptotic distributions of their position, velocity, and other relevant quantities. We summarize below our main findings.

\begin{itemize}[leftmargin=0pt]
\item As opposed to the position of the shock front, which becomes asymptotically deterministic, the position of the impurity remains a non-self-averaging random quantity. We argue that in the long-time limit, the typical impurity displacement depends on the shock front radius $R$ and the mean-free path in the initial state $\lambda$ as:
\begin{equation}
\label{R-imp}
R_\text{imp} = \lambda\left(\frac{R}{\lambda}\right)^{h_d}.
\end{equation}
In two-dimensions, this yields $R_{\text{imp}} \propto t^{2/5}$. We test this prediction in simulations, and a comparison with the theory is shown in \fig{numerical-test}(A).

\item The typical value of the impurity velocity is given by
\begin{equation}
V_\text{imp} = \sqrt{E}\,\left(\frac{a}{\lambda}\right)^\frac{d-1}{2}
    \left(\frac{\lambda}{R}\right)^{\omega_d},
    \quad\omega_d = \frac{d}{2}-\frac{d^2}{8 + 3 d^2}.
    \label{V-imp}
\end{equation}
Since ${a}/{\lambda} \sim \phi$, the volume fraction (see \eqref{phi:def} and \eqref{eq:mfp}), we can re-write \eqref{V-imp} as
\begin{equation}
\label{V-imp-phi}
V_\text{imp} = \sqrt{E}\,\phi^\frac{d-1}{2}\left(\frac{\lambda}{R}\right)^{\omega_d}.
\end{equation}
For the two-dimensional case, this gives $V_{\text{imp}}\propto t^{-2/5}$ and a comparison between theory and simulation is shown in \fig{numerical-test}(B). On the other hand, the hydrodynamic flow near the impurity is predominantly radial and follows the relation $u_{\text{imp}} \sim R_{\text{imp}}/t$. More precisely, we find
\begin{equation}
u_{\text{imp}} = \sqrt{E} \phi^\frac{d-1}{2}\left(\frac{\lambda}{R}\right)^{\frac{d}{2}+1-h_d}.
\end{equation}
For two-dimensions, this gives $u_{\text{imp}}\propto t^{-3/5}$. A comparison between theory and simulation is shown in \fig{numerical-test}(D).

\item We show that the typical number of collisions, $C_{\rm imp}$, experienced by the impurity scales as
\begin{equation}\label{C-imp}
C_\text{imp} = \left(\frac{R}{\lambda}\right)^\frac{8+2d^2}{8+3d^2}.
\end{equation}
For the two-dimensional case, this leads to $C_{\text{imp}}\propto t^{2/5}$. A comparison between theory and simulation is shown in \fig{numerical-test}(C).

\item As part of our heuristic derivation of some of the above results, we also show that in the core region, the characteristic  density $n_0$ and temperature $T_0$ are given by
\begin{equation}
\label{eq:coren}
n_0 = n_\infty \left(\frac{\lambda}{R}\right)^{d - 2\omega_d}
\end{equation}
and
\begin{equation}
\label{eq:coreT}
T_0 = E\left(\frac{a}{\lambda}\right)^{d-1}
    \left(\frac{\lambda}{R}\right)^{2\omega_d}.
\end{equation}
In two-dimensions, $n_0 \propto t^{-1/5}$ and $T_0 \propto t^{-4/5}$; see \fig{numerical-test}(E)-(F) for the comparison with simulation results.

\end{itemize}

The mean free path depends on the radius of spheres, a microscopic variable. Therefore, in contrast to the position of the shock wave, \eqref{R-shock}, predicted by ideal compressible hydrodynamics, the determination of the displacement of the impurity seemingly requires a microscopic framework. Indeed, the derivation of \eqref{R-imp} relies on the arguments used in kinetic theory \cite{fluid, Beijeren}. Remarkably, the length scale  $R_\text{imp}$ is identical to the one mentioned in \eqref{size-dim}, that arises in a more accurate hydrodynamic description of the flow in the core of the blast. Thus, the impurity remains inside the core. Note that the core-bulk boundary is not sharp, and what the previous line implies is that the impurity displacement and the core size have the same scaling with time.

The above formulas apply when $d\geq 2$, particularly in two and three dimensions. The one-dimensional setting is pathological due to the lack of hydrodynamic behavior: There is no mixing since two elastically colliding particles exchange the velocities and effectively pass through each other. In our setting, there is only one moving particle at any moment. One circumvents this pathology by allowing for distributed particle masses. The hydrodynamic behavior then emerges \cite{AKR,blast-1d,blast-1d-PF}. The core region was originally detected in one dimension \cite{blast-1d, blast-1d-PF}. The behavior of the impurity remains pathological since, in one dimension, the impurity is forever caged between the right and left neighbors. This caging property can be circumvented by postulating that the impurity colliding with a host particle passes with a certain probability $p$, while elastic collision occurs with probability $1-p$. This trick allows one to devise a meaningful version of the one-dimensional Lorentz gas \cite{Piasecki79, KR97, henk1982}. With such an amendment, the displacement of the impurity should be comparable with the size of the core region. Alternatively, one can consider a quasi-1D geometry where particles are allowed to overtake each other. Below, we always assume that $d\geq 2$ and employ the basic setting with spheres of equal masses undergoing elastic collisions.
\begin{figure*}
\centering
\includegraphics[width=0.32\linewidth]{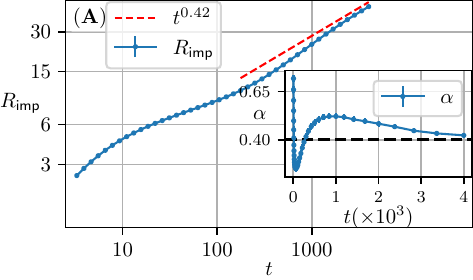}
\includegraphics[width=0.32\linewidth]{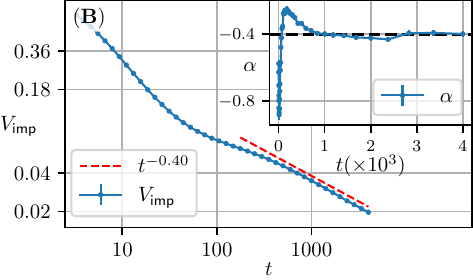}
\includegraphics[width=0.32\linewidth]{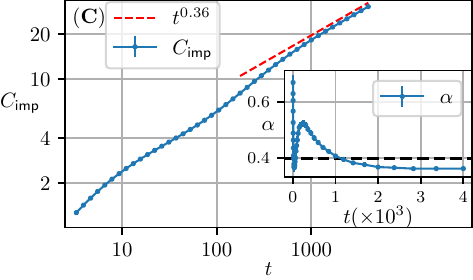}
\includegraphics[width=0.32\linewidth]{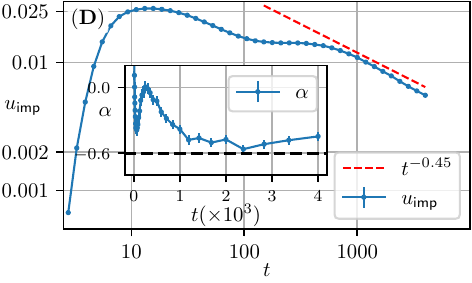}
\includegraphics[width=0.32\linewidth]{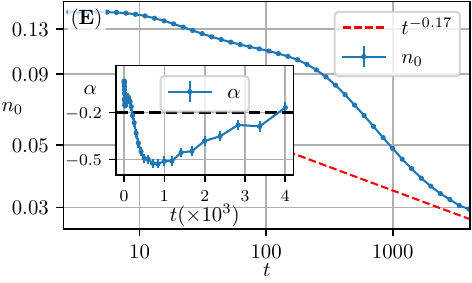}
\includegraphics[width=0.32\linewidth]{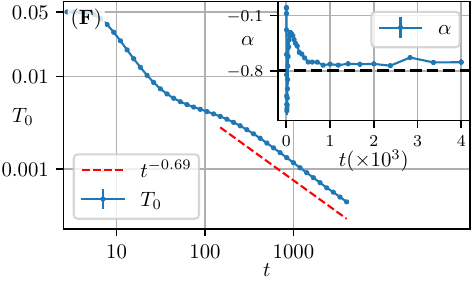}
\caption{
Time evolution of various impurity and fluid properties plotted on log-log scale in case of a $2$-dimensional hard-sphere gas:
(A) impurity displacement $R_{\text{imp}}$;
(B) impurity speed  $V_{\text{imp}}$;
(C) impurity collision count $C_{\text{imp}}$;
(D) fluid speed around the impurity $u_{\text{imp}}$;
(E) central density $n_0$;
(F) central temperature $T_0$.
All quantities are asymptotically expected to have power-law time dependence of the form $t^\alpha$ and the insets show the running temporal exponents plotted with time (see \sectn{test} for details). The dark dashed lines correspond to the theoretical predictions. All figures and their insets include error bars. However, the estimated statistical uncertainties are smaller than the symbol size and the line thickness used in the plots; as a result, the error bars are not visually discernible.}
\label{fig:numerical-test}
\end{figure*}

\section{Heuristic Derivations}
\label{sec:der}

The strategy of our heuristic derivation is that we first establish the properties in the core region where dissipation (viscosity and thermal conductivity) is important. Since we are primarily interested in scaling laws, throughout the derivation we will typically ignore many pre-factors. In particular, we find the typical size \eqref{size-dim} of the core region, and the density \eqref{eq:coren} and temperature \eqref{eq:coreT} in this region. Next, to justify the scaling laws \eqref{R-imp} and \eqref{V-imp-phi}--\eqref{C-imp}, we {\em assume} that in the long time limit, the impurity is in the core region and demonstrate self-consistency, thus confirming {\em a posteriori} that the impurity is inside the core. The derivation of Eqs.~\eqref{R-imp} and \eqref{V-imp-phi}--\eqref{C-imp} then follows from arguments based on elementary kinetic theory. The growth law $H\propto R^{h_d}$ for the characteristic size of the core region was established in \cite{blast-2d}. We now derive the more accurate result \eqref{size-dim}. 

To estimate the size of the core region where dissipative effects matter, it suffices to keep heat conduction and ignore viscous dissipation \cite{blast-2d, blast-1d, blast-1d-PF}. In this situation, the entropy equation reads
\begin{equation}
\label{entropy-heat}
\rho T(\partial_t + u \partial_r)s
= r^{-(d-1)}\,\partial_r (r^{d-1}\kappa  \partial_r T),
\end{equation}
where $u$ is the hydrodynamic radial velocity,
\begin{equation}
\label{entropy-def}
s = \frac{d}{2}\,\ln\!\left(\frac{T}{\rho^{2/d}}\right)
\end{equation}
the entropy per unit mass (assuming the ideal gas form), and $\kappa$ the coefficient of thermal conductivity. According to kinetic theory \cite{fluid,Beijeren,Landau-K}, for the dilute hard-sphere gas, the coefficient of thermal conductivity is proportional to the thermal velocity $\sqrt{T/m}$ and inversely proportional to the cross-section area $\sigma\sim a^{d-1}$. It is convenient to write $\kappa$ as
\begin{equation}\label{kappa-HS}
\kappa = n_\infty \lambda \sqrt{T}
\end{equation}
where we ignore the numerical pre-factor. Using
\eqref{entropy-def} we recast \eqref{entropy-heat} into
\begin{equation}
\label{heat}
\rho^{1+2/d}\,(\partial_t + u \partial_r)\frac{T}{\rho^{2/d}}
= \frac{1}{r^{d-1}}\, \partial_r \big(r^{d-1} \kappa\,\partial_r T\big)
\end{equation}
To estimate the radius $H$ of the core, defined as the region where the heat transfer is comparable with the terms on the left-hand side, we first insert  \eqref{kappa-HS} into \eqref{heat}. Approximating $\partial_t T \sim T/t, \partial_r T \sim T/H$ and $u \sim H/t$   we get the following relation between the size of the core $H$, time $t$, and the core density, $\rho$:
\begin{equation}\label{H-est}
\rho \sim n_\infty \lambda\, \sqrt{T}\, \frac{t}{H^2}.
\end{equation}
To find estimates of $\rho$ and $T$, we use the TvNS solution which is expected to be valid for $r \gtrsim H$. From the exact solution outside the core  \cite{blast-2d} one finds the asymptotic behaviors valid when $H \lesssim r\ll R$:
\begin{subequations}
\begin{align}
\label{r-est}
\frac{\rho}{n_\infty} &\sim\left(\frac{r}{R}\right)^\frac{d^2}{2}\\
\label{T-est}
T&\sim\left(\frac{R}{t}\right)^2 \left(\frac{r}{R}\right)^{-\frac{d^2}{2}}
\end{align}
\end{subequations}
The divergence of the temperature when $r\to 0$ is physically impossible. The asymptotic behaviors \eqref{r-est}--\eqref{T-est} are valid for $r\gtrsim H$. Setting $r=H$ in \eqref{r-est}--\eqref{T-est} and inserting into \eqref{H-est} we arrive at \eqref{size-dim}.

The density and temperature in the core region are comparable to the density $n_0$ and the temperature $T_0$ at the center of the explosion. The density $n_0$ is estimated by setting $r=H$ in \eqref{r-est}:
\begin{equation}
\label{n0}
n_0 = n_\infty \left(\frac{\lambda}{R}\right)^{d-2\omega_d}
\end{equation}
The temperature $T_0$ is similarly determined by setting $r=H$ in
\eqref{T-est}:
\begin{equation}
\label{T0}
T_0 = E\,\left(\frac{a}{\lambda}\right)^{d-1}\,
        \left(\frac{\lambda}{R}\right)^{2\omega_d}
\end{equation}

The typical speed of the impurity is essentially the thermal speed $\sqrt{T_0}$ in the core. Using $V_\text{imp} = \sqrt{T_0}$ and \eqref{T0} we arrive at \eqref{V-imp} and \eqref{V-imp-phi}.

To derive the typical displacement, we note that the impurity exhibits a diffusive motion, albeit with a decreasing self-diffusion coefficient. According to kinetic theory, the self-diffusion coefficient in the hard-sphere gas is the product of the mean-free path by thermal velocity \cite{fluid,KRB,Beijeren}. Thus, $D_0=\lambda_0 \sqrt{T_0} =\lambda_0 V_\text{imp}$. We use the mean-free path $\lambda_0=\lambda n_\infty/n_0$ inside the core, so the self-diffusion coefficient reads
\begin{equation}
\label{D:imp}
D_0=\lambda\, \frac{n_\infty}{n_0}\,V_\text{imp}
=\lambda\sqrt{E}\,\phi^\frac{d-1}{2}\, \left(\frac{\lambda}{R}\right)^{3\omega_d-d}
\end{equation}
The typical displacement of the impurity is found from $R_\text{imp} \sim\sqrt{D_0 t}$ which in conjunction with Eqs.~\eqref{D:imp}, \eqref{phi:def}, and \eqref{R-shock} lead to the announced estimate \eqref{R-imp}.

Thus {\em assuming} that the impurity stays within the core region, we have demonstrated the consistency by showing that the typical displacement of the impurity grows as the size of the core region. The impurity certainly stays behind the shock wave, but assuming that it explores the $r\leq R(t)$ ball and repeating the above line of reasoning gives $R_\text{imp}$ that grows slower than $R\propto t^{2/(d+2)}$. This failure has led us to the $R_\text{imp}\sim H$ guess that turned out to be self-consistent.

The hydrodynamic flow in the vicinity of impurity is radial, $u_\text{imp}\sim R_\text{imp}/t$. Thus,
\begin{equation}
\label{u-imp-phi}
u_\text{imp} = \sqrt{E}\,\phi^\frac{d-1}{2}\left(\frac{\lambda}{R}\right)^{\frac{d}{2} + 1 - h_d}
\end{equation}

Focusing on the time dependence, we have
\begin{subequations}
\label{RVu-HS}
\begin{align}
R_\text{imp} &\propto t^{\frac{2}{d+2}-\frac{8}{(d+2)(8+3d^2)}}\\
V_\text{imp} &\propto t^{- \frac{d}{d+2}+\frac{2d^2}{(d+2)(8+3d^2)}}\\
u_\text{imp} &\propto t^{-\frac{d}{d+2}-\frac{8}{(d+2)(8+3d^2)}}
\end{align}
\end{subequations}
Note that $u_\text{imp}(t)$ decays faster than $V_\text{imp}(t)$. This justifies using $V_\text{imp}$ and neglecting the hydrodynamic radial velocity in estimating the self-diffusion coefficient.

To determine the number of collisions experienced by the impurity, we use the Boltzmann cylinder argument \cite{fluid,Beijeren} and estimate the time interval $\tau = (n_0 a^{d-1} V_\text{imp})^{-1}$ between collisions at time $t$. The total number of collisions is estimated from $C_\text{imp}=t/\tau$. Using Eq.~\eqref{V-imp-phi} and Eq.~\eqref{n0}, we obtain
\begin{equation}
\label{C-imp-long}
C_\text{imp}
= \left(\frac{R}{\lambda}\right)^\frac{8+2d^2}{8+3d^2}
\propto t^\frac{4(4+d^2)}{(2+d)(8+3d^2)}
\end{equation}

Table~\ref{table:scaling} summarizes the scaling laws in two and three dimensions for the gas of hard spheres. Only the dependence on time is shown.

\begin{table}[h]
\begin{tabular}{| c | c | c | c | c | c | c |}
\hline
~$d$~& ~$C_\text{imp}$~ &   ~$R_\text{imp}$~      & $V_\text{imp}$   & $u_\text{imp}$     & $n_0$                  & $T_0$   \\
\hline
2 &  $t^{2/5}$                    & $t^{2/5}$                      & $t^{-2/5}$          &   $t^{-3/5}$           &     $t^{-1/5}$         & $t^{-4/5}$      \\
\hline
3 &  $t^{52/175}$              & $t^{62/175}$               & $t^{-87/175}$     &  $t^{-113/175}$     & ~$t^{-36/175}$~  & $t^{-174/175}$ \\
\hline
\end{tabular}
\caption{The temporal growth law for the number of collisions experienced by the impurity, Eq.~\eqref{C-imp}, and the typical displacement of the impurity, Eq.~\eqref{R-imp}; the latter quantity grows similarly to the size of the core region, Eq.~\eqref{size-dim}. Two following columns present the decay laws for the typical speed of the impurity, Eq.~\eqref{V-imp-phi}, and the hydrodynamic radial velocity in the vicinity of the impurity, Eq.~\eqref{u-imp-phi}. The two last columns give the density, Eq.~\eqref{n0}, and the temperature, Eq.~\eqref{T0}, at the center of the explosion.  We show the asymptotic behaviors in the physically relevant cases of two and three dimensions.}
\label{table:scaling}
\end{table}

\section{Position and momentum distributions}
\label{sec:PQ}

The collision dynamics is deterministic, but the outcome depends on the initial locations of the static spheres. Hence, we can treat the position and velocity of the impurity as random variables. The corresponding distributions $P(\bm{r},t)$ and $Q(\bm{v},t)$ cannot be separately studied since the position and velocity are correlated. For instance, if the impurity has a larger than average speed, it is expected to be further away from its initial position. Thus, only the joint distribution $\Pi(\bm{r},\bm{v},t)$ provides an adequate (and comprehensive) description. The joint distribution encodes the individual distributions:
\begin{equation*}
 P(\bm{r},t) =\int d\bm{v}\,\Pi(\bm{r},\bm{v},t), 
 \quad
 Q(\bm{v},t) =\int d\bm{r}\,\Pi(\bm{r},\bm{v},t)
\end{equation*}

Let us ignore correlations and try to probe the position distribution using a physically appealing but uncontrolled approximation. Employing a diffusion equation with a time-dependent diffusion coefficient is a naive approach. Indeed, the hydrodynamic radial velocity leads to $H\sim R_\text{imp}$ drift comparable with displacement generated by diffusion. A more consistent but still uncontrolled approximation is based on a convection-diffusion equation,

\begin{equation}\label{eq:imp-conv-diff}
(\partial_t + u_{\rm imp}\partial_r)P = D_0\nabla^2 P.
\end{equation}
For two-dimensional hard discs gas, we have $u_{\rm imp}\sim t^{-3/5}$ while $D_0\sim t^{-1/5}$. Focusing on time dependence, ignoring numerical factors, and taking into account the spherical symmetry, we get a convection-diffusion equation, which in two-dimensions is given by:
\begin{equation}\label{eq:Prt-2d}
\big(\partial_t+t^{-\frac{3}{5}}\partial_r\big)P
= t^{-\frac{1}{5}}\big(\partial_r^2+\tfrac{1}{r}\partial_r\big)P.
\end{equation}
We seek the solution of \eq{Prt-2d} in the scaling form
\begin{equation}\label{eq:Prt-2d-scal}
P(r,t) = t^{-\frac{4}{5}}\,\mathcal{P}(\eta), \qquad \eta = t^{-\frac{2}{5}}r.
\end{equation}
The normalization requirement becomes
\begin{equation}\label{eq:2d-norm}
\int_0^\infty dr\,2\pi r\,P(r,t) = 2\pi \int_0^\infty d\eta\,\eta \mathcal{P}(\eta) = 1.
\end{equation}
Plugging \eq{Prt-2d-scal} into \eq{Prt-2d} we obtain a second-order ordinary differential equation,
\begin{equation}\label{eq:ode_prob_2d}
-\frac{4}{5}\,\mathcal{P} -\frac{2}{5}\,\eta\, \mathcal{P}'
= \mathcal{P}''+\frac{1}{\eta}\,\mathcal{P}',
\end{equation}
where prime denotes the derivative with respect to $\eta$. Multiplying \eq{ode_prob_2d} by $\eta$ and integrating yields the first-order ordinary differential equation,
\begin{equation}\label{eq:ode_prob_2d_2}
-\frac{2}{5}\,\eta^2\mathcal{P} = \eta \mathcal{P}'.
\end{equation}
The solution to \eq{ode_prob_2d_2} satisfying the normalization requirement \eq{2d-norm} is
\begin{equation}\label{eq:2d-scal}
\mathcal{P} = \left(\frac{1}{5\pi}\right) \exp\!\left[-\frac{1}{5}\,\eta^2\right].
\end{equation}

The same approach can be repeated in three dimensions as well to obtain similar result in $3$-dimensions. In 3D, the analog of \eq{Prt-2d} is,
\begin{equation}\label{eq:Prt-3d}
\big(\partial_t+t^{-\frac{113}{175}}\partial_r\big)P
= t^{-\frac{51}{175}}\big(\partial_r^2+\tfrac{2}{r}\partial_r\big)P.
\end{equation}
while the analog of the scaling form \eq{Prt-2d-scal} is,
\begin{equation}\label{eq:Prt-3d-scal}
P(r,t)
= t^{-\frac{186}{175}}\,\mathcal{P}(\eta), \qquad \eta
= t^{-\frac{62}{175}}r.
\end{equation}
Plugging \eq{Prt-3d-scal} into \eq{Prt-3d} and solving the resulting ODE gives,
\begin{equation}\label{3d-scal}
\mathcal{P}
= \left(\frac{31}{175\pi}\right)^\frac{3}{2}
\exp\!\left[-\frac{31}{175}\,\eta^2\right].
\end{equation}

The omitted numerical factors do not affect the Gaussian form. Thus, the convection-diffusion approach leads to the Gaussian solution in two-dimensions. The extension to any $d\geq 2$ is straightforward and gives the same result:
\begin{equation}\label{Prt-d}
P(\bm{r},t)
= \frac{1}{(2\pi R_\text{imp}^2)^{d/2}}\,
\exp\!\left[-\frac{r^2}{2R_\text{imp}^2}\right],
\end{equation}
with $R_\text{imp} = C_d \lambda(R/\lambda)^{h_d}$. The numerical factor $C_d$ cannot be fixed in the realm of the above approach. We emphasize that have not justified the convection-diffusion approach, so the Gaussian form \eqref{Prt-d} of the position distribution is questionable.

A closed-form equation governing the evolution of the velocity distribution $Q(\bm{v},t)$, even a description based on uncontrolled approximation like the convection-diffusion equation for the position distribution, is unknown. The simplest guess is that the velocity distribution is asymptotically Gaussian. We do not have rational arguments in favor of this conjecture.

In contrast to the position and velocity distributions, the joint distribution satisfies a closed Lorentz-Boltzmann equation. In three dimensions
\begin{subequations}\label{LB-full}
\begin{equation}\label{LB}
(\partial_t + \bm{v}\cdot\nabla)\Pi(\bm{r},\bm{v},t)
= 4\pi a^2[\Gamma_+ - \Gamma_-].
\end{equation}
The terms on the right-hand side of Eq.~\eqref{LB} account for collisions of the impurity with host spheres:
\begin{align}\label{gain}
\Gamma_+
&=\int d\bm{e}\int d{\bf w}\,|{\bf w}-{\bf v}|
F(\bm{r},\bm{w}^*,t)\Pi(\bm{r},\bm{v}^*,t),\\
\label{loss}
\Gamma_-
&=\Pi(\bm{r},\bm{v},t)\int d{\bf w}\,|{\bf w}-{\bf v}| F(\bm{r},\bm{w},t)
\end{align}.
\end{subequations}
In the gain term \eqref{gain}, we denote by $\bm{v}^*$ the pre-collision velocity of the impurity and by $\bm{w}^*$ the pre-collision velocity of a host sphere colliding with impurity
\begin{equation}
\bm{v}^*=\bm{v}+\bm{e}[\bm{e}\cdot(\bm{v}-\bm{w})],
\quad
\bm{w}^*=\bm{w}-\bm{e}[\bm{e}\cdot(\bm{v}-\bm{w})],
\end{equation}
where $\bm{e}\in \mathbb{S}^2$ is the deflection angle and the integration measure is normalized: $\int d\bm{e}=1$. Due to local thermodynamic equilibrium, the joint distribution of the host spheres is locally Maxwell
\begin{equation*}
F(\bm{r},\bm{w},t)
= \frac{\rho(r,t)}{[2\pi T(r,t)]^{3/2}}\,
\exp\!\left\{-\frac{m|\bm{w}-u(r,t)|^2}{2T(r,t)}\right\}.
\end{equation*}
The density $\rho(r,t)$, temperature $T(r,t)$, and radial velocity $u(r,t)$ satisfy the Navier-Stokes equations in the core region. In the long time limit, these scalar fields acquire the scaling form depending on a single variable $r/H$. The scaled density, temperature, and radial velocity satisfy nonlinear coupled ordinary differential equations that have not been solved in the core region.

The governing Eq.~\eqref{LB-full} is a linear integro-differential equation that seems analytically intractable. Indeed, the distribution $F(\bm{r},\bm{w},t)$ describing host spheres contains the density, temperature, and radial velocity. Analytical expressions of these fields inside the core region are unknown, so the integral in Eq.~\eqref{loss} is already unknown.

Another challenge is to analyze the number of collisions experienced by the impurity beyond the typical value, Eq.~\eqref{C-imp}, estimated via arguments based on kinetic theory. Even for the dilute hard-sphere gas in thermal equilibrium, the collisional statistics of the tagged particle is remarkably subtle \cite{Lue05, Trizac08a, Trizac08b, Paik14}.


\section{Numerical verification}
\label{sec:test}

To test the predictions about the characteristics of impurities, we simulated hard discs in two-dimensions using an event-driven molecular dynamics. The system is prepared by randomly distributing discs, while ensuring the no overlap condition, within a simulation box of dimensions $L \times L$, at a fixed specified number density $n_\infty=N/L^2$ and volume fraction ($\phi=\pi a^2 n_\infty$). We chose four discs closest to the origin and gave them random momenta, ensuring that the total energy is exactly $E$ and the total momentum vanishes. The simulation proceeds by calculating the times of all possible future collisions between discs and with the walls, then evolving the system to the earliest collision event. At each event, the velocities of the colliding particles are updated based on the conservation of momentum and energy. Between collisions, the discs move linearly with constant velocity. In all simulations performed, we used density $n_\infty=0.15$, the radius of particles $a=1/2$, and the mass of particles $m=1$.

The system contained $N=40000$ particles, and the event-driven molecular dynamics simulations ran until $t=4000$. For computing the properties of impurities, we averaged over $\nR$ randomly chosen initial conditions, while for computing bulk and surface quantities, we used $\nRsurf$ initial conditions. For any observable, say $A$, we expect power law behavior, $A\sim t^\alpha$, in the long time limit, while at finite times, there will be corrections. Hence, in our numerics, we estimate the temporal evolution of this exponent for various quantities of interest. We recorded observables (averaged over realizations) at exponentially separated times in powers of $2^{1/4}$. We then obtained the running time exponents from the data at successive times (namely $\alpha(t)= 4[\ln A(2^{1/4}t)-\ln A(t)]/\ln 2$). In \tbl{results}, we also show the errors in computing these exponents, $\delta\alpha$.

{\renewcommand{\arraystretch}{1.5}
\begin{table}[h]
\begin{tabular}{|c|c|c|c|c|c|c|}
\hline
& $R_{imp}$ & $C_{imp}$ & $V_{imp}$ & $u_{imp}$
& $T_{0}$ & $n_{0}$ \\
\hline
$\alpha_T$ & $2/5$ & $2/5$ & $-2/5$ & $-3/5$
& $-4/5$ & $-1/5$\\
\hline
$\alpha_S$ & $\aRimp$ & $\aCimp$ & $\aVimp$ & $\auimp$
& $\aTo$ & $\ano$\\
\hline
$\delta\alpha_S$ & $\daRimp$ & $\daCimp$ & $\daVimp$ & $\dauimp$
& $\daTo$ & $\dano$\\
\hline
\end{tabular}
\caption{The table shows the comparison between theoretically computed and
numerically estimated exponents $\alpha$ for time $(t)$ dependencies of the
form $t^\alpha$ for the quantities of interests.
$\alpha_{T}$ and $\alpha_{S}$ stands for Theory and Simulation.
 $\delta\alpha$ is the error in estimation of exponent $\alpha$ due to
statistical error in the respective quantity.} 
\label{tbl:results}
\end{table}
}

{\bf Impurity properties}:
Let ${\bf r}(t)$ be the displacement, at time $t$, of any of the four impurity particles from its initial position and ${\bf v}(t)$ be its velocity. We define the typical impurity displacement and velocity as:
\begin{align}\label{eq:Rimp_comp}
R_{\rm imp}(t) = \langle {\bf r}^2/2 \rangle^{1/2},~~~~
V_{\rm imp}(t) = \langle {\bf v}^2/2 \rangle^{1/2},
\end{align}
where the average is over $n_R=3\times10^4$ different initial conditions and over the four impurities. We define $C_{imp}(t)$ as the mean number of collisions that any impurity undergoes up to time $t$. The comparison between the kinetic predictions and the simulation results are shown in \fig{numerical-test}$(A)$-$(C)$.

To compute the fluid speed, $u_{imp}$, around the impurity particle, we find the average flow velocity of all the particles within a fixed radius around the impurity and compute its magnitude. The fixed radius is chosen so that, at $t=0$, there are  $100$ other gas particles within that distance. Again, we take an average over the four impurities and $n_R$ realizations. The comparison of $u_{\text{imp}}$ between theory and simulations is shown in \fig{numerical-test}$(D)$.

Finally we numerically computed the radial distribution function, $p_{\rm imp}(r,t)$,  of the impurity's displacement at time $t$ and compared it with the  predicted form from Eq.~\ref{Prt-d},
\begin{equation}
\label{eq:pr}
p(r,t)=2\pi rP({\bf r},t)
=\frac{r}{R_{\rm imp}^2}\exp\left(-\frac{r^2}{2R_{\rm imp}^2}\right).
\end{equation}
In \fig{Rimp} we see that there is close agreement between the simulation results and the predicted form.

\begin{figure}
\centering
\includegraphics[width=\linewidth]{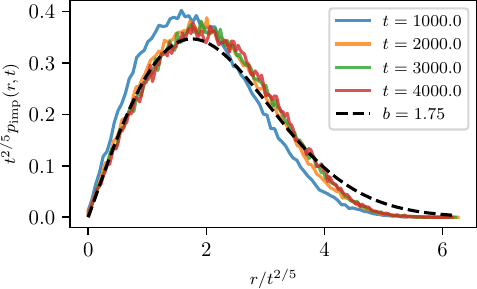}
\caption{
Scaled radial probability distribution function $(p_{\rm imp})$ for impurity (obtained from simulation) plotted at different times. The dashed curve in the figure is plotted by fitting \eq{pr}) with $R_{\text{imp}}=b~t^{2/5}$ against the PDF obtained from simulation data at $t=4000$. The parameter $b$ obtained by fitting is $1.75$ which is close  to $b=1.71$, obtained directly from computation of $R_{imp}/t^{2/5}$ at $t=4000$.}
\label{fig:Rimp}
\end{figure}

{\bf Core properties}:
For the computation of the quantities in the blast core, namely central density $n_0$ and central temperature $T_0$, we consider a circular region that includes $100$ particles at $t=0$. Then, we count the total energy and the number of particles in this circular region. Dividing the energy by the number of particles inside the circular region gives the central temperature, $T_0$, while dividing the number of particles by the area of the central region gives the density,  $n_0$. The results for  $n_0$ and $T_0$ are shown in \fig{numerical-test}(E)-(F).

A summary of our results for various exponent values and their comparison with the analytic prediction reported is given in  \tbl{results}. The reported exponent values correspond to the last two recorded times.


\section{Summary}
\label{sec:disc}

The large-scale motion generated by kicking an impurity in a homogeneous cold gas is equivalent to the hydrodynamic behavior in the earlier stage of the atomic explosion when the pressure behind the shock wave greatly exceeds the atmospheric pressure. The hydrodynamic behavior is tractable in this regime. The pressure ahead of the shock vanishes in the zero-temperature gas, so the shock remains infinitely strong, and the analytical solution applies forever.

We analyzed the evolution of the impurity. We probed the typical displacement, the average speed, and the average number of collisions experienced by the impurity. The chief observation is that the impurity remains inside the core region where the standard description of the infinitely strong blast based on the Euler equations becomes inaccurate, and one should rely on the Navier-Stokes equations accounting for dissipative processes.

The impurity in the cold homogeneous hard-sphere gas exhibits algebraic scaling behaviors with amusing rational exponents (Table~\ref{table:scaling}). Our derivations of these scaling laws rely on arguments based on elementary kinetic theory along with inputs from hydrodynamics both at the Euler and dissipative scales. In dimension $d=2$, we performed molecular dynamics simulations to test our heuristic predictions and find reasonable agreement. We believe that simulations at longer times will lead to better agreement.

The linear Lorentz-Boltzmann equation describing the evolution of the joint position-velocity distribution of the impurity provides the theoretical framework for the comprehensive analysis of the impurity. Unfortunately, the Lorentz-Boltzmann equation appears analytically intractable. In addition to the mathematical challenges caused by the necessity to solve an integro-differential equation in an inhomogeneous and evolving background, we do not even have the analytical description of the background due to the lack of the analytical solution of the Navier-Stokes equations in the core region.

Impurities are generally different from the host particles. The disparity between the mass and radius of impurity and the mass and radii of host spheres may lead to interesting behaviors, especially when these disparities are significant. We leave this problem for the future. One can also explore the impurity problem in high spatial dimensions, $d\gg 1$, which may be more amenable to analytical treatment and perhaps sheds light on the behavior in the physically relevant three-dimensional case. Finally, we mention a different setting where the impurity is normally incident on the static gas occupying the half-space \cite{AKR}. The impurity, as well as every host sphere, is expected to be ejected backward into an initially empty half-space after a finite number of collisions. This assertion is intuitively clear, albeit proof could be challenging. So far, the splash problem has been quantitatively studied in one-dimension \cite{blast-splash, kumar2025splash, splash25}.

\section{Acknowledgement}
We thank Subhadip Chakraborti and Frank Smallenburg for useful discussions on numerical simulation techniques. AD and UK  acknowledge support from the Department of Atomic Energy, Government of India, under project No.~RTI4001. AD acknowledges the J.C. Bose Fellowship (JCB/2022/000014) of the Science and Engineering Research Board of the Department of Science and Technology, Government of India.

\bibliography{references-gas}

\end{document}